\documentclass[useAMS,usenatbib]{mnras}
\usepackage{graphicx}

\title[Does feedback help or hinder star formation?]{Does feedback help or hinder star formation? The effect of photoionisation on star formation in Giant Molecular Clouds}

\author[Shima, Tasker \& Habe]{Kazuhiro Shima$^1$\thanks{E-mail:shima@astro1.sci.hokudai.ac.jp}, Elizabeth J. Tasker$^{1,2}$, Asao Habe$^1$ \\
$^1$ Department of Physics, Faculty of Science, Hokkaido University, Kita 10 Nishi 8 Kita-ku, Sapporo 060-0810, Japan \\
$^2$ Institute of Space and Astronomical Science, Japan Aerospace Exploration Agency, Yoshinodai 3-1-1, Sagamihara, Kanagawa, Japan
}

\begin{document}
\date{}

\pagerange{\pageref{firstpage}--\pageref{lastpage}} \pubyear{2015}
\maketitle

\label{firstpage}

\begin{abstract}
  We investigated the effect of photoionising feedback inside turbulent star-forming clouds, comparing the resultant star formation in both idealised profiles and more realistic cloud structures drawn from a global galaxy simulation. We performed a series of numerical simulations which compared the effect of star formation alone, photoionisation and photoionisation plus supernovae feedback. In the idealised cloud, photoionisation suppresses gas fragmentation at early times, resulting in the formation of more massive stars and an increase in the star formation efficiency. At later times, the dispersal of the dense gas causes the radiative feedback effect to switch from positive to negative as the star formation efficiency drops. In the cloud extracted from the global simulation, the initial cloud is heavily fragmented prior to the stellar feedback beginning and is largely structurally unaffected by the late injection of radiation energy. The result is a suppression of the star formation. We conclude that the efficiency of feedback is heavily dependent on the gas structure, with negative feedback dominating when the density is high. 
\end{abstract}


\section{Introduction}

Stars form in cold cradles of molecular gas that are identified in observation and simulation as the Giant Molecular Clouds (GMCs). As dense gas cores within these clouds collapse into stars, they begin to emit heat into its surrounding gaseous nursery. This energy will affect the efficiency of the future star formation or may stop it completely if the cloud is dispersed. Such interplay between the parent cloud and its child stars therefore controls the star formation rate in the galaxy, but the determining variables are complex.

When a star adds energy to its environment it can do this through mechanical or radiative processes. In the former, momentum is imparted to the surrounded gas to drive winds that carry material away from the star. In the latter, the heated gas increases in pressure and blows bubbles of ionised HII gas \citep[see][for a detailed discussion of the different feedback mechanisms]{Krumholz2014}. If the stars are massive, then these effects can drive turbulence through the whole cloud.

What happens to the cloud next is a topic of intense debate. Sufficiently strong feedback must disrupt the cloud entirely, ending all prospects of future star formation in that particular structure \citep{Murray2011}. One notch down would see the cloud significantly damaged, delaying the on-set of a second generation of stars \citep{Meidt2015, Williamson2014, Tasker2015}. Alternatively, the effect of feedback might be positive. As feedback drives gas away from the star formation site, the outer edge of the resulting expanding shell can fragment into a population of triggered stars \citep{Whitworth1994, Wunsch2010, Koenig2012}. Heat from the newly forming stars can also increase the Jeans mass, reducing the gas fragmentation to produce more massive stars forming in place of a larger number of smaller objects \citep{Bate2009, Offner2009b, Urban2010}. 

Simulations have tried to determine which of these outcomes will dominate. The conclusion has been that the result depends not just on the type of feedback employed, but on the cloud itself. The mass and radius of a cloud controls its escape velocity; a value that affects the extent outflows can travel. When comparing the impact of HII regions and stellar winds on star formation, \citet{Dale2014} found that while HII regions played the dominant effect between the two mechanisms for small clouds, those with higher escape velocities suffered little impact. Similarly, more compact clouds will have a higher surface density, allowing radiation to be more efficiently trapped within the cloud where it can have a stronger effect. \citet{Krumholz2010} found that a high surface density allowed high accretion rates for the forming protostars, whose radiation from the accretion luminosity was then trapped in the dense cloud. The result was a rise in temperature that suppressed fragmentation to form more massive stars.

The differences do not stop with the global cloud properties. Clouds are not uniform gas distributions that form stars only within a dense central region. Rather, they are turbulent, irregular bodies that can harbour a large multiple of star formation sites \citep{Larson1981}. This means that the local conditions of the gas around the star formation site are a long way from being a homogeneous pool and these small-scale variations can play a key role. Heat that is deposited into dense gas will cool rapidly, reducing the region affected by that feedback. On the other hand, if stars form near pockets of low density gas, then the energy may have a much longer-range impact. Comparing observations of wind blown bubbles around stellar clusters with theoretical models, \citet{HarperClark2009} found that a non-homogeneous medium is needed to match observations, which allows energy to leak through the bubble shell. 

All this points to an efficiency for feedback that may come down less to the feedback itself and more to the internal structure of the cloud.

However, if the cloud structure is key, how can this be included self-consistently in feedback models? The properties of clouds have been shown to strongly depend on their galactic environment, with disc shear, grand design structure and neighbouring cloud interactions sculpting their evolution \citep{Fujimoto2014, Tasker2009, Meidt2015, Dobbs2015}. This makes a typical cloud's internal gas structure difficult to determine. Observations outside the Milky Way can now estimate the bulk properties of individual clouds, but not yet map their interior dynamics \citep{Hughes2013, DonovanMeyer2013}. Simulations suffer from similar problems, with those modelling the global galaxy disc creating self-consistent gas profiles but being unable to resolve the cloud interior, or alternatively following the gas inside the cloud but using an idealised initial set-up \citep{Federrath2014, Offner2009a}.

One way to tackle this is to try and bridge the two scales by using properties derived from a global simulation as the initial starting point for a smaller-scale model. Where this has been done, the importance of the gas structure has become clear. On parsec-scales within a single GMC clump whose structure is taken from numerical models, \citet{Rogers2013} looked at effect of wind-driven bubbles. They found that the variations in gas structure allowed hot, high speed gas to escape long low density channels, producing a strongly different effect from a uniform density environment. The energy from the final supernovae explosion largely escapes the now fractured gas shell. On slightly larger scales, \citet{ReyRaposo2015} compared the evolution and star formation of clouds extracted from a global galaxy simulation with those modelled as idealised turbulent spheres. While their models did not include feedback, they found that the differing velocity structures in the clouds produced very different evolutions. The sphere evolution was governed principally by gravitationally infall, while the extracted clouds had a more involved velocity structure from the galactic disc sheer. 

In this paper, we investigate the effect of photoionising feedback on the star formation within a giant molecular cloud. We look at two sets of cloud models. In the first, the cloud is modelled as an idealised turbulent sphere of gas. We compare the effects of star formation with no form of feedback with the changes when stars radiate and finally when old stars explode as supernovae, depositing thermal energy into the gas. In our second cloud model, the cloud is extracted from a global galaxy simulation. We compare the resulting evolution with star formation only and when including photoionisation. Section \S2 describes our numerical methods, section \S3 shows the results for the idealised turbulent cloud and \S4 the extracted simulated cloud. In Section \S5 we discuss our results and conclusions.

\begin{figure*}
  \includegraphics[width=168mm]{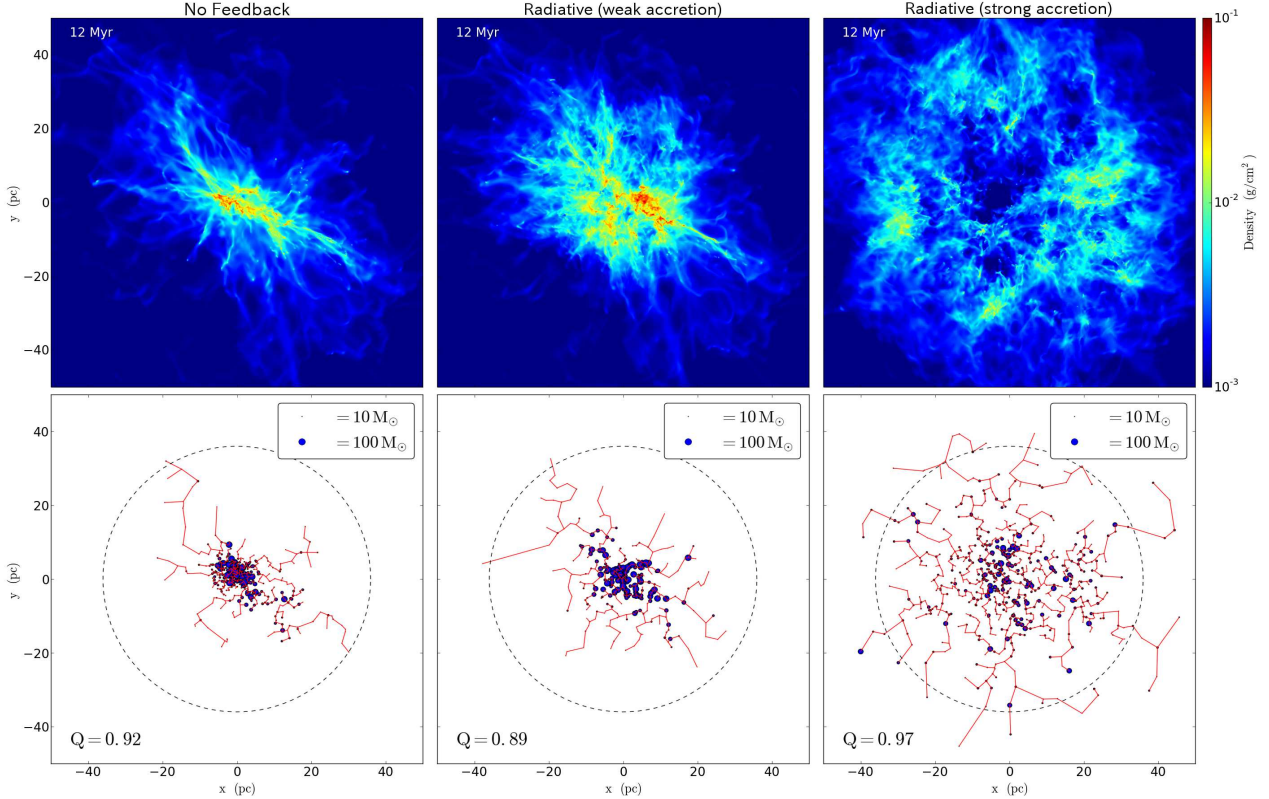}
  \caption{Comparison of the gas surface density (top row) for the idealised Bonnor-Ebert cloud when no feedback is included (left: run 1), feedback using the weak accretion model (centre: run 4) and feedback using the strong accretion model (right: run 3) after one free-fall time (12\,Myr). The average outward radial velocity for each cloud is $-3.4$\,km/s (collapsing), $1.7$\,km/s and $3.5$\,km/s respectively. The bottom row shows the star particle distribution corresponding to each image above. The blue circles are proportional to the star particle mass, while the red solid line shows the minimum spanning tree (see section~\ref{sec:q}) which connects all the points. The quantitative description of the fragmentation, $Q$, is shown in the bottom left corner of each panel. The black dashline marks the initial cloud radius.}
  \label{cf_projection}
\end{figure*}

\section{Numerical methods}

Our simulations were run with the adaptive mesh refinement (AMR) hydrodynamics code, {\it Enzo} \citep{Enzo}. The gas was evolved with self-gravity using a three-dimensional implementation of the {\it Zeus} hydro code \citep{Zeus}, where the main parameter, the artificial viscosity term, was set to its default value of 2.0. Cells were refined based on the baryon mass and the Jeans length, whereby four cells or greater must be shorter than one Jeans length. This criteria follows the \citet{Truelove1997} suggestion as the minimum needed to prevent spurious numerical fragmentation. At the maximum refinement level where the Jeans criteria inevitably must break, we introduce a pressure floor in the form of a polytrope where the adiabatic index, $\gamma = 2.0$. This halts the collapse at a finite density, preventing individual cells becoming unphysically massive. In practice, the pressure floor is rarely used as overdense cells form star particles.

{\it Enzo} follows nine atomic and molecular species, ${\rm H, H^+, He, He^{+}, He^{++}, e^{-}, H_2, H_2^+, H^-}$ for non-equilibrium cooling, and supplements this with metal cooling using data precomputed using the CLOUDY photoionisation software, assuming a solar hydrogen mass fraction and solar metallicity \citep{CLOUDY,Ferland1998}. The mean molecular weight is calculated from the species abundances, giving a value $\mu\sim$1.2. The radiative cooling then extends down to 10\,K, which is the temperature of the giant molecular clouds. 

For our idealised cloud simulations, the box size has a side of $200$\,pc, covered by a $128^3$ root grid and an additional two static meshed corresponding to a minimum $512^3$ resolution over the cloud. An additional three levels of adaptive refinement were included, with each static and adaptive mesh reducing the cell size by a factor of two. This gave a limited resolution of 0.05\,pc. We performed two additional high-resolution simulations (with and without photoionising radiation) that reduced the cell size by a further factor of two. For the simulation that used the extracted cloud, the box size was larger with side 500\,pc. This was on a $64^3$ root grid which corresponded to the global simulation's maximum resolution of 7.8\,pc. We added a further six levels of refinement to reach a limited resolution of 0.1\,pc. This is slightly larger than the idealised simulation case due to computational time. All gas denser than approximately $10$\,cm$^{-3}$ was resolved to at least $512^3$ (level 3).

\subsection{Star Formation}

Star particles form in the simulation when the gas flow converges into a (maximum refined) cell with a density greater than $10^4$\,cm$^{-3}$ and a temperature $\le 10$\,K. This threshold is user-defined, and we selected it to be the value at which star formation is observed to occur inside a GMC \citep{GMC1,GMC2, Kainulainen2014}. Since this density is significantly below stellar densities, the resulting particle is treated as a star cluster. When a star particle is formed, half the mass is removed from the cell to create the initial particle. The star particle's velocity is the average of the neighbouring cells to prevent a runaway phenomenon. In addition, to avoid too many radiation sources in a small region, any new stars forming within 3\,pc of an accreting star are merged. This mass accretion is from a sphere of cold ($T < 10^3$\,K) gas whose size is defined at each time step in two ways in our calculations. The first method is the default scheme used within {\it Enzo}. A sphere is found such that its average gas density corresponds to a dynamical time, $t_{\rm dyn} = 0.5$\,Myr; approximately one free-fall time for gas at our threshold density of $10^4$\,cm$^{-3}$. The second method uses the Bondi-Hoyle accretion radius (or a gravitational capture radius), defined as $R_{\rm BH} = 2GM/(v^2+c_s^2)$, where $G$ is the gravitational constant, $M$ is the star particle mass, $v$ is the relative velocity between the star particle and the accretion sphere gas, and $c_s$ is the sound speed. This radius indicates the region in which gas will be caught by the star's gravitational potential. 

The typical accretion sphere size of the second Bondi-Hoyle method is smaller than the free-fall time sphere, leading to a lower accretion rate and smaller stars. For these simulations, the typical size of the free-fall time sphere is approximately $\sim 4 \times \Delta x_{\rm min}$, where the smallest cell size, $\Delta x_{\rm min} = 0.05$. We can estimate the Bondi-Hoyle accretion radius for a star particle of 1\,M$_\odot$ and $c_s = v = 0.3$\,km/s to give $R_{\rm BH} \sim 0.05$\,pc. This corresponds to roughly one minimum cell size, $\Delta x_{\rm min}$. For higher particle velocities, the accretion radius will shrink and be rounded back to one cell, whereas for larger star particles of $100\,$M$_\odot$, the radius extends to $\sim 0.2$\,pc $\sim 4\times \Delta x_{\rm min}$. As most have mass less than $100\,$M$_\odot$, this gives a smaller typical accretion radius. 

Due to these differences in accretion radii, we define this as the `weak' accretion model as that where the Bondi-Hoyle accretion radius is used and the free-fall time sphere as the `strong' accretion model. The accretion continues to increase the star's mass for one dynamical time or until the particle hits $800$\,M$_\odot$. This star formation scheme is a slightly modified version of the cosmological star cluster method in \citet{popII}.

\subsection{Feedback}
\label{sec:feedback}

After the accretion has finished, the star particles emit ionising radiation using the adaptive ray tracing scheme implemented in {\it Enzo} that is described in \citet{ART1,ART2} and based on the HEALPix framework \citep{HEALPIX}. Each star particle has an ionising luminosity of $3\times 10^{46} {\rm ph/s/M_{\odot}}$ \citep{Schaerer2003}. This value assumes solar metallicity and a Salpeter initial mass function between $1 - 100$\,M$_\odot$. Within this range, the IMF gradient does not depend strongly on model choice, for example giving a similar distribution to the Chabrier IMF \citep{Chabrier2005}. The rays are assumed to be monochromatic with the mean ionising photon energy of 21.6\,eV. These values were adopted from the defaults in {\it Enzo}, which assumes each star particle represents a stellar cluster. 

Since we resolve down to the masses of individual stars (although not to stellar densities), our ionising luminosity is likely an overestimate. Our radiative feedback should therefore be considered as an upper limit. 

In one simulation, we also include thermal feedback from supernovae explosions.  After 4\,Myr, massive clusters with $M > 100$\,M$_\odot$ deposit thermal energy equal to $1\times 10^{49}$\,erg/M$_\odot$ into its surrounding cell. This is equivalent to one supernova per $100$\,M$_\odot$ depositing $\sim10^{51}$\,erg of energy; a frequency consistent with the Salpeter IMF for the cluster. Since supernova are actually distributed in time between 4 - 40\,Myr, the deposit of energy at the lower limit of 4\,Myr suggests this feedback rate is the upper limit \citep{Krumholz2014}. After 4\,Myr, the star effectively `dies' and stops radiating. In the simulations without supernovae feedback, star particles continue to radiate indefinitely. 

\section{Initial Conditions}

We consider two separate initial conditions in this paper. The first uses an idealised Bonnor-Ebert profile for the cloud, while the second extracts a cloud that formed in a global galaxy simulation. 

\subsection{Bonnor-Ebert cloud}
 
Our idealised cloud takes the density profile of a Bonnor-Ebert sphere \citep{Bonnor1956}; a hydrostatic isothermal self-gravitating sphere of gas that is confined by its external pressure. While such a profile is derived analytically, there is observational evidence of their existence in nature, such as the Bok Globule B68 \citep{Alves2001}. Our clouds slightly exceed the maximum stable mass for the Bonner-Ebert profile and therefore begin to collapse after the start of the simulation. The resulting cloud has a mass of $9.64\times 10^4$\,M$_\odot$, with an initial radius of 36.3\,pc. 

The cloud is given additional internal support from an initial injection of turbulence, produced by imposing a velocity field with power spectrum $v_k \propto k^{-4}$. This corresponds to the expected spectrum given by Larson for GMCs \citep{MacLow1998, Larson1981}. The turbulence slows the collapse as the gas cools and creates a filamentary structure of dense regions, instead of a centralised collapse. Since we did not want the cloud to be strongly distorted by its turbulence, we removed the lower order modes to avoid the large-scale perturbations. We also used an upper limit, corresponding to a maximum $k$-mode that was 1/10th of the number of cells across the cloud. This was to ensure adequate resolution of the included modes. This selection corresponded to  $6 < k < 19$ for the GMC. The turbulence amplitude was set by the Mach number, $\cal{M} \equiv {\rm \sigma_c} / {\rm c_s}$, where $\sigma_c$ is the velocity dispersion inside the cloud and $c_s$ is the sound speed. At the start of the simulation, the initial temperature is the Bonner-Ebert equilibrium temperature of 1200\,K  and $\cal{M}$ = 1. The cloud cools rapidly, leaving the turbulence to support the cloud. A summary of the cloud properties is shown in Table~\ref{be_simulation_data}.

This cloud was used in four simulations: (1) without feedback, (2) with the strong feedback from the free-fall time accretion radius, (3) with the weak feedback using the Bondi-Hoyle accretion radius and (4) with the addition of supernovae feedback. In all cases, the evolution time for the run was one free-fall time, corresponding to 12\,Myr.

\begin{table}
 \begin{minipage}{140mm}
  \caption{Idealised (Bonnor-Ebert) cloud parameters}
  \label{be_simulation_data}
  \begin{tabular}{@{}lll@{}}
  \hline
 $\Delta x_{\rm min}$         & 0.05      & pc \\
 $R_{\rm c}$              & 36.3      & pc\\
 $M_{\rm c}$                & $9.65\times 10^4$    & M$_{\odot}$\\
 $T_{\rm c}$         & 1200      & K\\
 $\bar{\rho_c}$     & 19.5 & atoms/cc\\
 $t_{\rm c, ff}$            & 12.0      & Myr\\
 $\sigma_c$ & 3.77      & km/s\\
  \hline
  \end{tabular}
 \end{minipage}
\end{table}

\begin{table}
 \begin{minipage}{140mm}
  \caption{Extracted cloud parameters}
  \label{ec_simulation_data}
  \begin{tabular}{@{}lll@{}}
  \hline
 $\Delta x_{\rm start}$          & 7.8        & pc \\
 $\Delta x_{\rm min}$           & 0.1        & pc \\
 Main clump radius, $R_c$  & 26.2       & pc \\
 Main clump mass, $M_c$    & 4.4e+6    & M$_{\odot}$ \\
 $M_{\rm total}$    & 1.4e+7     & M$_{\odot}$\\
  \hline
  \end{tabular}
 \end{minipage}
\end{table}

\begin{table*}
  \caption{Simulations performed. Columns show run number, the initial conditions (idealised Bonnor-Ebert or extracted global simulation cloud), the method for calculating the accretion radius, inclusion of photoionising radiation, whether the star radiates continuously from formation (on) or stops after 4\,Myr (off), inclusion of supernova and minimum cell size.}
   \label{run_table}
   \begin{tabular}{lllllll}
   \hline
   Run  & IC & Accretion Type & Photoionisation & Cont. radiation & Supernova & $\Delta x_{\rm min}$ \\
   \hline	 
  1  &  BE cloud &   strong & off   &  n/a & off & $0.05$\,pc\\
  2  &  BE cloud &   weak   & off   & n/a   &  off & $0.05$\,pc \\
  3  &  BE cloud &   strong & on     & on   &  off & $0.05$\,pc\\
  4  &  BE cloud &   weak   & on   & on   &  off & $0.05$\,pc\\
  5  &  BE cloud &   strong & on   & off &  on & $0.05$\,pc \\
  6  &  BE cloud &   strong & off   & n/a &  off & $0.025$\,pc \\
  7  &  BE cloud &   strong & on   & on &  off & $0.025$\,pc \\
  8  &  Sim. Extract &   strong  & off  & n/a & off & $0.1$\,pc\\
  9  &  Sim. Extract &   strong  & on  & on  & off & $0.1$\,pc\\
   \hline
   \end{tabular}
\end{table*}

\subsection{Extracted cloud}
\label{sec:ic_simcloud}

Our second set of initial conditions extracts a cloud from a global galaxy disc simulation. The global simulation was also run using {\it Enzo} and is described in detail in \citet{Benincasa2013}. The galaxy has the form of a Milky Way-type disc, with a flat rotation curve of 200\,km/s. A rotating frame of reference exists at a radius of 6\,kpc, making gas at that radius stationary with respect to the grid, while gas at smaller and larger radii flows in opposite directions. This minimises the artificial support from the Cartesian mesh. The clouds are identified as connected cells with density over $100$\,cm$^{-3}$ (details in \citet{Tasker2009}) and their properties are found to agree well with those observed in nearby disc galaxies. Clouds from this simulation were extracted to be used as initial conditions for more detailed star formation calculations and can be found online at {\tt http://www.physics.mcmaster.ca/mcclouds/}. In this online catalogue, we used cloud with tag number 1149.

The extracted region is 500\,pc across and contains a total gas mass of $1.4\times 10^7$\,M$_\odot$ . Within the box, there is the central body of the cloud which has two clumps of high density gas, surrounded by a lower density network of tidal tails from these clumps interacting. The larger of the two clumps has a mass of $4.4\times 10^6$\,M$_\odot$ and radius of 26.2\,pc while the smaller one is roughly half as massive, with $2.0\times 10^6$\,M$_\odot$ and 22.5\,pc in radius. Unlike the idealised cloud case, this is clearly not a passive environment, but a fragmented and highly interactive location.

The \citet{Benincasa2013} simulation did not contain any star formation or feedback. While this had the advantage of resolving to higher resolutions more easily without having to negotiate large particle sizes, it did mean the gas had become very dense. Such a large pool of over dense gas would immediately turn to stars, producing an unphysical starburst and injection of feedback energy. To avoid this problem, we first evolved the gas for the crossing time of the box (calculated as the box size divided by the maximum velocity in the central clumps) equal to 6\,Myr, and increased both the resolution and cooling to alter the minimum temperature from 300\,K (used in the global model) to 10\,K. We used a non-accreting star formation method, which ate away at the dense gas, converting it into particles. After these 6\,Myr, we then turned on our free-fall / strong accretion star formation model. Due to the heavy computation time, we performed this run with the strong accretion and without supernovae. The details of the simulation set-up are outlined in Table~\ref{ec_simulation_data}.

A summary of the performed runs in this paper is given in Table~\ref{run_table}.

\section{Results: Idealised Cloud}

\begin{figure}
  \includegraphics[width=84mm]{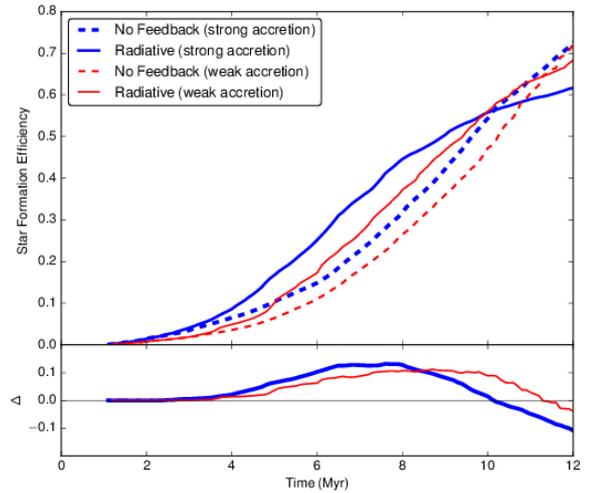}
  \caption{Comparison of the star-formation efficiency between runs with no feedback (dashed lines) and runs including radiative feedback (solid lines). Models that use the weak accretion model are in red, while the results from the strong accretion model are in blue. These are runs 1 - 4 in Table~\ref{run_table}. The star formation efficiency is defined as  $M_{\rm star}(t) / M_{\rm cloud}(t=0)$. $\Delta$ shows the difference between solid and dashed lines, with a +/- value indicating the positive/negative effect of feedback.}
  \label{cf_efficiency}
\end{figure}

\begin{figure*}
	\includegraphics[width=170mm]{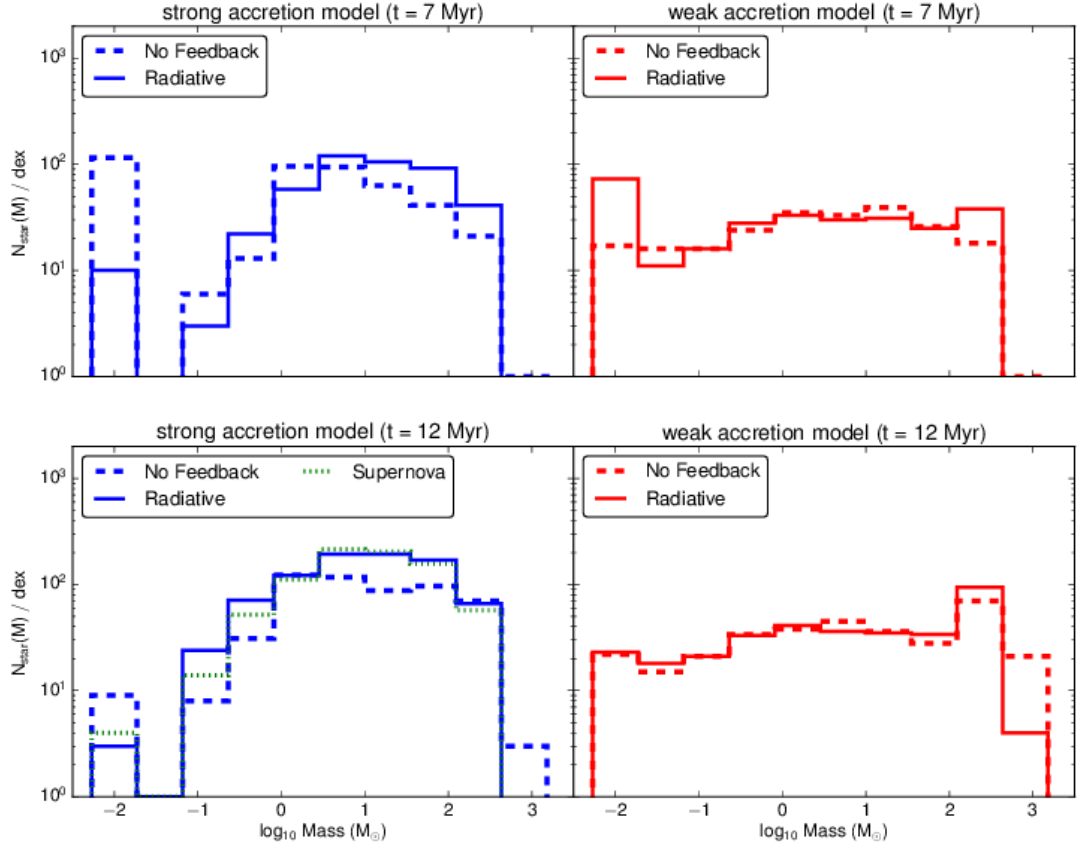}
	\caption{Comparison of the stellar mass distribution at $7.0$\,Myr (top row) when the effect of feedback is positive (boosting star formation) and at $12.0$\,Myr (bottom row) when it is negative. Dashed lines show simulations without feedback, while solid lines are for radiative photoionising feedback. (Runs 1 - 4 in Table~\ref{run_table}.) Green dotted line shows the simulation that also includes supernovae feedback (Run 5). The left-hand panels in blue are for the strong accretion model, while the right-hand panels in red show results for the weak accretion model.}
	\label{cf_massfunction}
\end{figure*}

\begin{figure}
	\includegraphics[width=84mm]{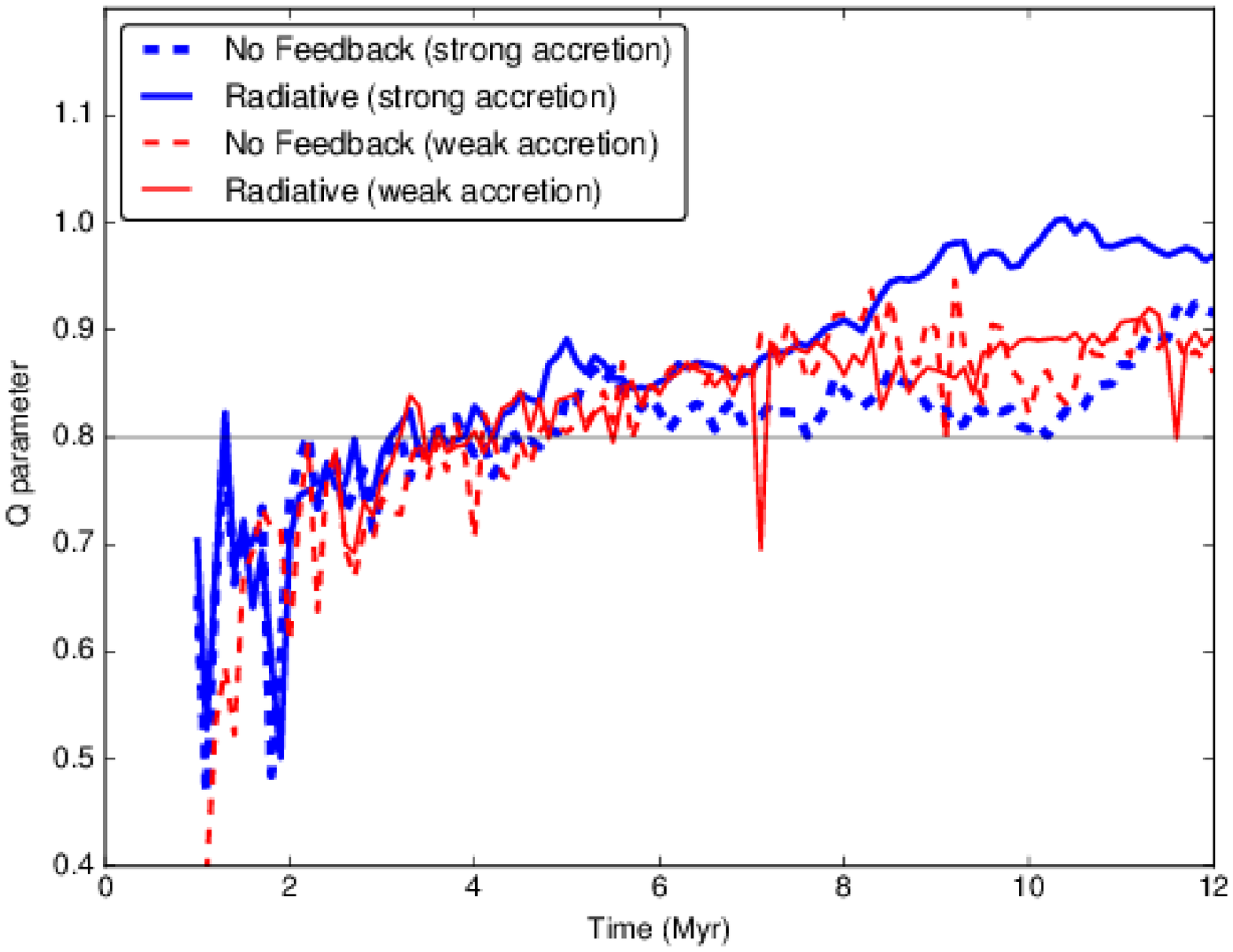}
	\caption{Comparison of the Q-parameter for each of the first four runs with the idealised cloud. A value of $Q \simeq 0.8$ indicates a constant volume density of stars through the cloud, while higher and lower values suggest a concentrated and more fractal distribution, respectively.}
	\label{cf_Qparameter}
\end{figure}

\subsection{Cloud Morphology}

The top row of images in Figure~\ref{cf_projection} shows the gas surface density for simulations using the idealised Bonnor-Ebert sphere after one free-fall time (12\,Myr) (Runs 1, 4, and 3 in Table~\ref{run_table}). The left-most panel shows the simulation without any feedback, with stars formed using the strong accretion model. The middle panel includes radiative feedback using the weak accretion model for the star formation while the right panel includes radiative feedback with the strong accretion model. We performed an additional run without feedback using the weak accretion model, but the results matched that in the left-hand panel.

In all three cases, the initial turbulence in the gas causes it to form a filamentary and fragmented structure. As the gas cools, the turbulence decays and self-gravity dominates. The gas begins to collapse, increasing in density until the highest density regions reach our star-formation threshold. In the absence of any stellar feedback, the formation of stars does not halt the collapse, which continues under the cloud's own gravity. At the time shown for the non-feedback case, the gas is collapsing inwards with an average radial rate of -3.4\,km/s. 

When radiative transfer is included in the simulation, the forming star particles generate thermal pressure that counters the collapse. With the weak accretion model, the gas is left expanding at an average radial rate of 1.7\,km/s. This increases to 3.5\,km/s for the stronger accretion model, removing all dense gas from the central region. These results strongly suggest that star-forming clouds are heavily impacted by their stellar-feedback and can even be disrupted.

\subsection{Star Distribution}
\label{sec:q}

The bottom row in Figure~\ref{cf_projection} shows the projected position of the star particles that formed in each cloud. The size of each blue circle marking a star particle is proportional to the particle mass and the black dashed line marks the radius of the initial cloud. The particles are connected with red lines that represent a `minimum spanning tree' whereby all points are linked such that the total length of the connecting lines is minimised and there are no closed loops. Using this structure, a quantitative value ($Q$) was developed by \citet{Q} for defining how stars are distributed within a cluster or (in our case) cloud. This $Q$-parameter is defined as:

\begin{equation}
  Q = \frac{\langle l \rangle}{\langle s \rangle}
\end{equation}

\noindent where $\langle l \rangle$ is the mean length in the minimum spanning tree between star particles and $\langle s \rangle$ is the average separation between any two particles. If the stars are distributed evenly through the cloud to produce a uniform volume density, then $Q \simeq 0.8$. Values higher than this indicate that the stars are more centrally concentrated, while lower values imply a fragmented, fractal distribution of the stellar population. 

In all three cases for the idealised cloud, the $Q$ value is higher than the uniform case, showing that the star formation activity is largest in the cloud centre. This is unsurprising, since the gas in that region begins with the highest density, reaching the star formation threshold to produce the first population of stars prior to any feedback. In the case with no feedback, the stellar population remains localised in the centre of the collapsing cloud, with only a small number of stars forming along filamentary structures perpendicular to the direction of collapse. When feedback is included in the weak accretion model, the star population spreads outwards as the global collapse is reversed to allow dense gas to form further from the centre. With feedback using strong accretion model, stars form much further from the central region as the gas more rapidly expands. A small number of star particles are even found at radii beyond the original cloud edge. These stars actually formed within the cloud boundary, but escaped outwards. Comparing with the above panel showing the gas surface density, it can be seen that the central stars must be older, as there is now very little dense gas in that region. Despite this significantly more distributed population, the $Q$ value remains high, showing that there is still a steady gradient in the star population density towards the centre of the cloud, rather than multiple individual sites of high star formation activity. The evolution of the $Q$ value over the cloud lifetime will be considered in section~\ref{sec:q-evol}.

\subsection{Star Formation Efficiency}

How effectively the cloud converts its gas into stars is measured by the star formation efficiency (SFE), defined as the total stellar mass divided by the initial gas mass:
\begin{equation}
	\epsilon(t) = \frac{M_{\rm star}(t)}{M_{\rm cloud}(t=0)}
\end{equation}

\noindent This is shown in Figure~\ref{cf_efficiency} for times throughout the simulation. The SFE for the non-feedback simulations is shown by the dashed thin red lines for the weak accretion case and dashed thicker blue lines for the strong accretion case. For when radiative feedback is included, the line is solid with red and blue once again showing the weak and strong accretion cases, respectively. The difference between the feedback and non-feedback runs for each type of accretion is shown by the value $\Delta$ in the bottom panel of that plot, where $\Delta$ is simply:
\begin{equation}
	\Delta(t) = \epsilon_{\rm Feedback}(t) - \epsilon_{\rm No Feedback}(t)
\end{equation}
\noindent giving a positive value when the feedback promotes star formation and a negative value when the star formation is suppressed. 

At roughly 2\,Myr, the first star formed in the simulation finishes its accretion and begins to emit ionising radiation during the feedback runs. Shortly after this, all four runs begin to deviate to launch into a different SFE history. The inclusion of feedback initially promotes the production of stars, raising the SFE above the non-feedback runs in both the weak and strong accretion models. This is reflected in the $\Delta$ value, which climbs during the first half of the simulation. The origin of this increased SFE from feedback could come from a number of sources. Star formation could be triggered in the edges of swept-up expanding shells of gas, producing a small but numerous population. Alternatively, the freshly heated gas could prevent fragmentation, forming a larger reservoir for newly formed stars to accrete to create more stellar mass than that from multiple smaller star particles. In the next section, we will see it is this second option that promotes the SFE. The stronger accretion model also forms larger stars, giving a higher SFE than the weaker accretion model for both the feedback and non-feedback runs.  

Just after half-way through the simulation, the $\Delta$ value for the strong accretion case turns over and begins to drop. This is followed at around 10\,Myr by the weak accretion case. The positive effect of feedback to boost star production drops until it becomes negative, and its presence suppresses the SFE compared to the non-feedback simulations. At this point, the solid lines dropping below the dashed in the upper SFE plot. This reversal in the effect of the feedback is due to the dispersal of the dense gas. As the cloud continues to expand, the gas that has not yet collapsed into stars is spread over a wider area. Without feedback, this lower density gas can continue to collapse into a late stage of star formation, but with the outward force of feedback, it is permanently dispersed. The switch between positive and negative feedback occurs first in the stronger accretion model, since the gas is being dispersed more rapidly by the larger stars producing stronger radiation.  

Despite suppression from the feedback, at the end of the simulation the SFE for the whole cloud is very high, varying between 60 - 70\%. This corresponds to a star formation rate per free-fall time of ${\rm SFR_{ff}} \sim 0.5$. By contrast, observations of GMCs suggest values of a few percent \citep{Krumholz2007}. Our higher numbers stem from the gravitational potential of the cloud overtaking the internal kinetic energy as the turbulence decays and is not sufficiently driven by the internal feedback. This suggests observed clouds may be only locally collapsing and globally supported by externally-driven turbulence or possibly magnetic fields \citep{Federrath2015}.

\subsection{The Stellar Mass Distribution}

The range of masses of the star particles formed in the idealised cloud simulations are shown in Figure~\ref{cf_massfunction}. In all four panels, the dashed line shows the non-feedback runs, while the solid line is for feedback. Blue lines (left-hand plots) show the star particle masses when the strong accretion is used, while the red lines (right-hand plots) are for the weak accretion. The top two panels show the stellar mass distribution half-way through the simulation at 7.0\,Myr, when the effect of radiative feedback is positive and boosts the star formation. The bottom two panels shown the result at the end of the simulation, 12\,Myr, where the feedback now suppresses the star formation compared to the non-feedback runs.

At 7\,Myr, the strong accretion model has a median star particle mass of 1.0\,M$_\odot$. The total number of star particles formed in all simulations is very similar at roughly 400 particles, but their masses vary. The mass profile is more peaked for the strong accretion model than when the weak accretion model is used, reflecting the ability to form larger star particles more easily during the accretion phase. For the strong accretion run, the effect of feedback is to reduce the number of small star particles and form instead, larger stars. This suggests the impact of feedback here is not primarily to trigger a population of star particles in the expanding shells of gas, but to suppress fragmentation. The outer layers around a newly forming star are thrown outwards by the radiative feedback and heated. The surrounding gas therefore increases and warms, stopping its fragmentation but making it available to be accreted by nearby star particles which gain in mass.

\begin{figure}
  \includegraphics[width=84mm]{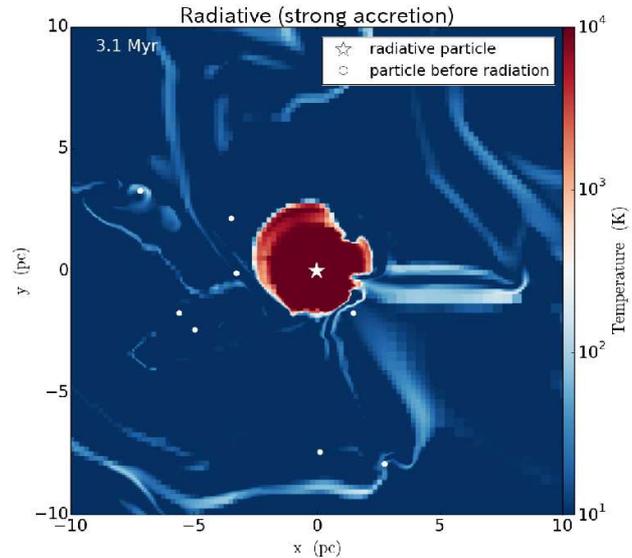}
  \caption{ The temperature distribution (slice) at 3.1\,Myr for the strong accretion run, centred around a star particle 0.6\,Myr after it began to emit radiation. Neighbouring star particles within 10\,pc are also shown as white dots.}
  \label{cf_temperature}
\end{figure}

To confirm this situation, we measured the average accretion rate for the star particles with and without feedback. The accretion rate increased when feedback was used by roughly a factor of 2.0, confirming that the feedback impact is to provide more gas to build larger stars. This is different from the triggered star formation scenario, where feedback drives expanding shells of gas which fragment into new stars in a `collect and collapse' sceanario. Figure~\ref{cf_temperature} shows a typical example of a hot expanding shell within our cloud at 3.1\,Myr after the start of the simulation and 0.6\,Myr after the beginning of radiation feedback. The radiating star particle (denoted as a central white star) has a mass greater than $> 800$\,M$_\odot$. Smaller white dots in the image show surrounding star particles that sit within 10\,pc of the central star and are currently accreting mass. There is no evidence around this shell or others in our simulation of collect and collapse star formation and we see no evidence of an elevated star formation around expanding shells. This is contrary to the smaller clouds explored by \citet{Walch2013}, who find that a central star can trigger further star formation around an expanding shell.  In our simulations, the turbulent gas produces a complex structure of filaments within which the star is born and this makes it hard to create a well defined shell wall.

In the weaker accretion case, the feedback also increased the number of the most massive stars ($> 100$\,M$_\odot$) through the same mechanism. However, there is also a boost in the quantity of the smallest stars around $0.01$\,M$_\odot$. It could be that triggered star formation is occurring in the weaker accretion case as the slower expansion of the cloud promotes the dense gas shells. This is hard to confirm visually and these small particles are merged with close-by larger neighbours, causing them to disappear by the later bottom-right panel. The number of small stars does decrease between the two time steps in all runs. This is due to accretion increasing the mass of the stars and a smaller number of new stars being born at later times. 

By the end of the simulation, the stellar mass distribution (bottom panels) shows that feedback has created generally more massive stars for the strong accretion case. There is a population of very massive ($> 1000$\,M$_\odot$) particles for the non-feedback run, which correspond to the final gravitational collapse of the cloud. This situation is mirrored to a smaller extend in the weak accretion run, with feedback producing a slight excess of star particles with $M > 100$\,M$_\odot$.

\subsection{Q-parameter}
\label{sec:q-evol}

The distribution of star particles through the cloud is shown by the evolution of the $Q$-parameter in Figure~\ref{cf_Qparameter}. In all runs, $Q$ increases over time, indicating that a concentrated profile develops, with the largest number of star particles in the cloud centre. While the strong accretion feedback produces the highest number of star particles away from the core, as shown in Figure~\ref{cf_projection}, it has the steepest gradient through the cloud, landing it a high $Q$ value. Without feedback, most of the stars are in a smaller region around the centre, producing a nearly uniform volume density with $Q$ value slightly over 0.8. The weak accretion model is insufficient to change this significantly. These values are similar to those found from observations of GMCs, which display a range of $Q$-parameters from 0.7 - 1 in different clusters \citep{Q-parameter}.

\subsection{Supernovae Feedback}

We performed one additional simulation using the strong accretion model with radiative feedback, where a massive star's radiation stops after 4\,Myr, concluding with a supernovae thermal energy injection, as described in section~\ref{sec:feedback}. The difference the addition supernovae feedback makes is very small, since relatively few massive stars were formed in the simulation and the ionising radiation has already cleared away most of the gas. This can be seen in the lower right-panel of Figure~\ref{cf_massfunction}. The stellar mass distribution is only weakly affected by the supernovae. At the lower mass end of the final stars formed ($< 1$\,M$_\odot$), the supernovae run forms less stars than when the radiation continues steadily but slightly more of the smallest population of stars at $0.01$\,M$_\odot$. This implies that a sudden thermal injection is less effective at changing the fragmentation and accretion rate than continuous radiation energy.

\begin{figure*}
  \includegraphics[width=140mm]{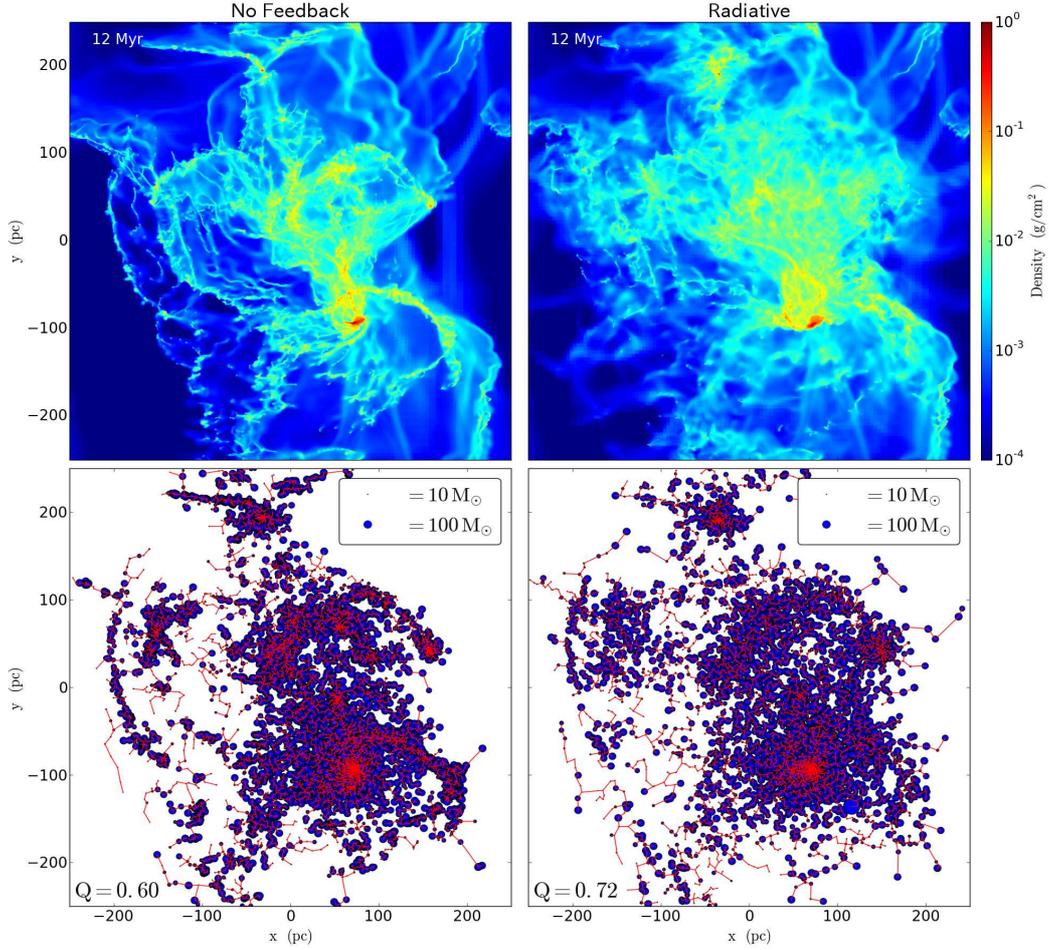}
  \caption{Comparison of the face-on gas surface density (top row) and the star particle distribution for the cloud extracted from the global model at 12\,Myr. The size of the blue points in the bottom panel is proportional to the star particle mass and red lines show the minimum spanning tree (see section~\ref{sec:q}). Left panels show the simulation without stellar feedback, while the right-panel includes radiative feedback.}
  \label{ec_projection}
\end{figure*}

\begin{figure*}
  \includegraphics[width=140mm]{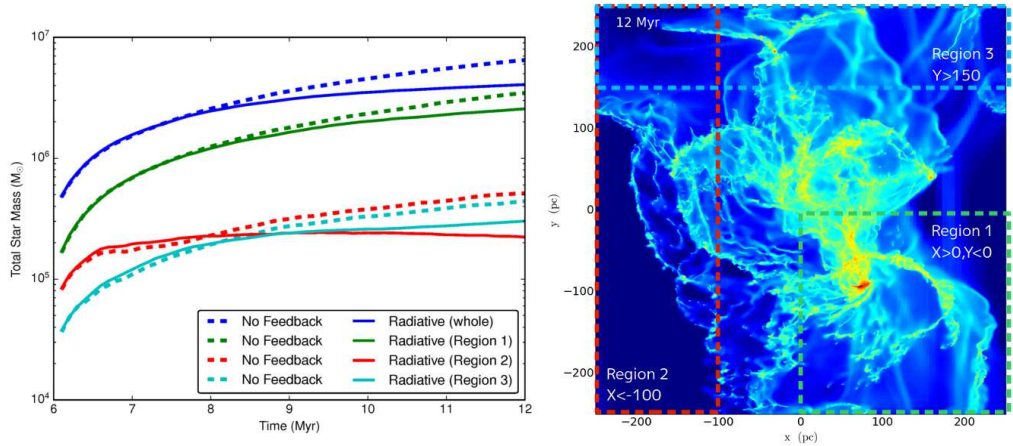}
  \caption{The evolution of the total star mass in the radiative feedback (solid lines) and non-feedback (dashed) simulations for different regions within the globally simulated cloud. Line colour corresponds to coloured box region, with the green region 1 containing the most high density gas, while regions 2 and 3 (red and blue) contain less dense material. The dark blue lines show the result for the whole box.}
  \label{ec_totalmass}
\end{figure*}

\begin{figure}
  \includegraphics[width=84mm]{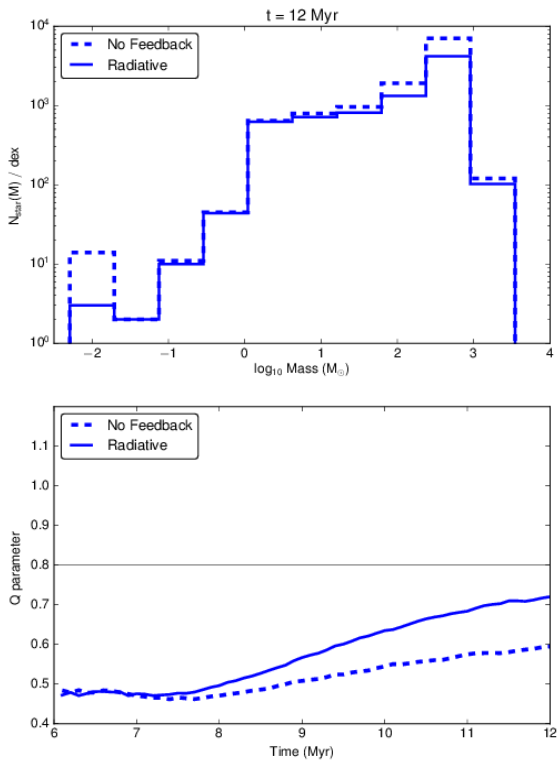}
  \caption{The stellar mass distribution at 12\,Myr and $Q$-parameter for the globally simulated cloud for both non-feedback (dashed) and feedback runs for one crossing time after the star formation begins.}
  \label{ec_massfunction_and_Q}
\end{figure}

\section{Results: Globally Simulated Cloud}

We now change from looking at the effect of feedback on the idealised Bonnor-Ebert cloud structure, to that of a cloud formed in a global galaxy simulation, as described in section~\ref{sec:ic_simcloud}, and listed as runs 8 and 9 in Table~\ref{run_table}.

This is a significantly different initial condition for the cloud. Formed in a dynamic environment feeling the effects of sheer and the gravitational pull of nearby clouds, the cloud here is not in equilibrium. Rather, it has already fragmented to produce a varied density structure. Below, we examine the impact of photoionisation on the cloud's late evolution.

\subsection{Cloud Morphology}

The morphology of the cloud after one crossing time (at 12\,Myr) is shown in Figure~\ref{ec_projection}. The panels show disc face-on projections of the gas surface density (top row) and the star particle positions and their minimum spanning tree (bottom). The left-hand side is for the run with star formation but no stellar feedback, while the right-hand images shown the effect of including photoionisation. 

An immediate difference between this cloud and the idealised cases in the previous section is that the radiative feedback is making a much smaller impact on the cloud morphology. There is no evidence that the cloud is globally disrupted by the feedback. Rather, the changes appear on more local scales. The central dense region survives the injection of ionising radiation, but increases in radius. Surrounding pockets of dense gas also appear more diffuse and in the lower density filaments at the box edge, the feedback has disrupted their structure.

This is shown quantitatively by the star particle distribution and the $Q$-parameter in the lower panels. Without feedback, the stellar structure gives $Q = 0.6$, pointing to a fractal clustering of stars, rather than a uniform distribution or dominant centre. This agrees with the multiple small sites of dense (red) gas in the gas surface density above. Adding feedback increases  $Q$ to 0.72, implying that these centres have been partially disrupted to produce a more uniform distribution of stars.

\subsection{Total Stellar Mass}

Due to the simulated cloud containing a wide variety of environments, it is more helpful to look at the impact of feedback on different regions. Figure~\ref{ec_totalmass} shows the time evolution of the total stellar mass for both the non-feedback and feedback runs in the whole simulation box (dark blue) and in three different sub-regions. The regions are marked on the surface density image in the right-hand panel of the same figure. Region 1 (green lines) contains the densest clump in the box with an average gas density of 400\,cm$^{-3}$. Region 2 (red) and Region 3 (cyan) contain more low density gas, with the average in Region 2 of 230\,cm$^{-3}$ and in Region 3, 270\,cm$^{-3}$. Region 2 is very fractal, with a small $Q$ ($\sim 0.4$) and plenty of filamentary structure. Region 3 has a number of smaller star-forming clumps and a less fractal $Q$ value at 0.6. As it is no longer clear in the simulated cloud exactly where the cloud edge is, the SFE is a less helpful quantity so we instead focused on the total star mass. The time evolution begins with the formation of the first radiating star particle, where the accretion radius corresponds to our strong radiative feedback mode. 

In the densest Region 1, the addition of feedback has no effect on the star production for the first two million years. After that time, the two runs begin to deviate, with the feedback suppressing the star formation in the clumps. For the lower density Regions 2 and 3, the effect differs. Like the idealised cloud, the initial impact of feedback is to promote star formation, causing the radiative simulation to have a slightly higher stellar mass over the first 3 - 4\,Myrs. After that time, the feedback acts to suppress the star formation, making a larger difference to the gas than in the higher density Region 1.

This difference underscores the importance of the initial state of the gas. Radiative feedback in our simulations can only promote star formation if the gas is less dense. Within collapsing cores, its action is to lower the rate of star formation by providing a pressure to counterbalance the collapse. The pressure in this situation is not enough to overwhelm the self-gravity, but it can slow the production of star-forming gas. 

Considering the gas in the entire box, the SFE reaches $\sim$50$\%$ when feedback is not includes and decreases to $\sim$30$\%$ with feedback, agreeing with the result that star formation is overall suppressed in this denser environment.

\subsection{Stellar Mass Distribution and Q-parameter}

The distribution of stellar masses in the globally simulated cloud is shown in the top panel of Figure~\ref{ec_massfunction_and_Q}, with the dervied $Q$-parameter in the lower panel. Due to the larger amount of material available, gas in the simulated cloud collapses to form very massive stars. The density in these regions is too great to be disrupted by the feedback, which acts to slightly suppress the formation of stars with $M > 1$\,M$_\odot$. Where smaller stars are forming, the feedback has a stronger effect, causing fewer stars with $M < 0.1$\,M$_\odot$ to form. These lower mass stars are likely to be forming in lower density regions such as Region 2 and 3 in Figure~\ref{ec_totalmass}, where the gas is much less fragmented and feedback can have a significant effect in diffusing the dense clumps.

The $Q$-parameter for the cloud's stellar population steadily rises over the course of the simulation. This suggests that the star formation is initially very fractal, forming in multiple pockets of over-dense gas. With feedback unable to break up the dense cores, the gas steadily collapses, moving towards a more uniform distribution of stars. Feedback promotes this process. By suppressing star formation in the densest regions, it allows the star particle distribution to even out as lower density gas begins to collapse. 

Observations indicate that the $Q$-parameter tends to be higher than $\sim 0.7$, suggesting that this breaking of fractal structure by feedback is occurring in GMCs. Simulations performed by \citet{Dale2012b} indicate that this process is dependent on the gas density. In higher density regions, the feedback acted to raise $Q$, whereas in lower density gas it had the reverse effect. \citet{Dale2012b} accounted for this difference by the denser gas more successfully forming `collect and collapse' shells of star formation. Our simulations do not show strong evidence of this mechanism, but the reduction of star formation in the dense cores still leads to more distributed star formation and a higher $Q$.

\section{Discussion}

\subsection{Feedback: Positive or Negative?}

The impact of radiative feedback differed strongly between our two cloud types: the idealised Bonner-Ebert sphere and the cloud extracted from the global simulation. The main difference between these two models was density. The idealised cloud initially had a smooth density profile, with an average value three orders of magnitude below our star formation threshold (see Table~\ref{be_simulation_data}). The extracted cloud meanwhile, had evolved without forming stars in its global environment. It therefore has multiple regions that have already collapse to high density before the feedback was allowed to act. The difference produced the change between positive and negative feedback effects.

Both our weak and strong accretion models with feedback were effective in low density gas. As stars began to form, the feedback ejected the outer layers of gas and heated them, preventing further fragmentation and allowing neighbouring star-forming regions to accrete more effectively. The result was to increase the mass of the newly formed stars, creating a positive feedback effect on the total star formation. This continued until the star-forming regions became very dense. The feedback was then no longer able to eject hot gas into the regions surrounding the new star and could only slow the collapse of the dense gas. This reduced the star formation rate, producing a negative effect on star production. 

In the extracted cloud, much of the gas was already at this later collapsed phase. The feedback there could only slow the collapse and reduce the star formation, but it could not disperse gas to increase accretion on neighbouring stars. This meant that the overall effect of the feedback was negative, suppressing the star formation. The only exception to this was in the low density regions of the simulation box, where a small positive effect could be seen on the gas that was newly collapsing. 

This result agrees with findings by \cite{DaleMII}, who noted that very massive clouds are largely unaffected by feedback due to their high escape velocities preventing gas escaping. We note the same effect on a small scale, with feedback in star-forming clumps producing a negative effect due to the feedback being unable to escape and affect the surrounding medium.
ex
\subsection{The Effect of Resolution}

\begin{figure}
  \includegraphics[width=84mm]{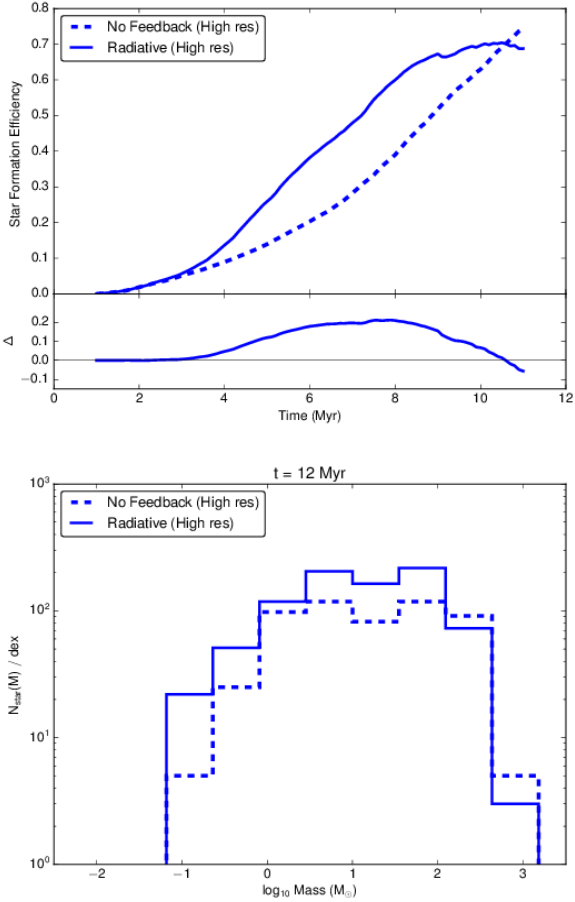}
\caption{The SFE evolution (top) and the stellar mass distribution at 12.0\,Myr for the idealised cloud at a factor of two higher resolution. The dashed line shows the simulation without feedback and the solid line is when radiative feedback is included. $\Delta$ in the top figure is the difference between the two runs, with a +/- value indicating a positive/negative effect from the feedback. Comparison with the main run in Figure~\ref{cf_efficiency} and Figure~\ref{cf_massfunction} shows that results are robust to resolution effects.}
  \label{cf_refine6}
\end{figure}

We performed two extra simulations at a higher resolution with $\Delta x = 0.025$\,pc, for idealised cloud using the strong accretion model, with and without radiative feedback. The trends observed, including the initial positive effect of the feedback changing to a negative impact, were unchanged. Figure~\ref{cf_refine6} shows the evolution of SFE for these runs and the stellar mass distribution at the end of the simulation, which is a close match to Figure~\ref{cf_efficiency} and the left-bottom panel of Figure~\ref{cf_massfunction}. This result therefore seems to be robust to resolution effects. 

\subsection{The Effect of Dust}

One effect that we did not include in our calculations is that of dust. Dust grains can absorb photons in addition to the gas, potentially resulting in a smaller expansion of feedback-driven bubbles from radiation pressure. About 25\% of the total photons may be absorbed by the dust assuming a Milky Way dust-to-gas ratio and then later re-emitted at infrared wavelengths \citep{McKee1997}. The impact of this process is difficult to estimate. Previous research suggests that the expansion of HII regions are primarily governed by thermal pressure except in dense starburst environments \citep{Krumholz2009}. It is likely true that if the dust density follows that of the gas, the impact will remain strongly dependent on the cloud structure. A more detailed model is required to investigate this further.

\section{Conclusions}

We investigated the effect of photoionisation on two different cloud structures: an idealised Bonnor-Ebert sphere that was initially stable and a complex cloud structure extracted from a global galaxy simulation. Our results are as follows:

\begin{enumerate}
  \item photoionisation can both promote star formation (positive feedback) and suppress it (negative feedback). Which occurs depends on the density of the gas. High density regions that are undergoing gravitational collapse could not be dispersed by our feedback. Instead, the collapse was slowed to reduce the star formation rate. On the other hand, lower density regions could have their outer layers blown out by the radiative feedback. In this case, the surrounding medium increases in density and temperature to become a pool for accretion material. Nearby star-forming regions increased in mass to form larger stars. 

\item The $Q$-parameter is a quantitative way of measuring the distribution of stars. Simple profiles like our idealised cloud move towards centrally concentrated profiles, even in the presence of feedback. The cloud formed in a global simulation had a more complex density structure, forming initially a more fractal distribution of stars that moved towards uniform as the gas collapsed.

\item Our simulation did not show obvious triggering in expanding shells from the feedback. However, the complex structure of filaments within the cloud makes this hard to detect. Instead, we find feedback primarily increased the number of more massive stars formed due to bolstering the accretion rate.

\item The addition of a thermal feedback supernovae term did not made a significant difference to the star formation. The supernovae exploded late in the star's lifetime after the ionising radiation had removed any surrounding dense gas.

\item We tried two different models for star formation accretion, using different accretion radii. While the results were broadly the same, feedback was significantly more effective when the average star mass was larger. While not surprising, this emphasises the sensitivity of the results to the forming stellar population.

\end{enumerate}

Our ultimate conclusion is that the effect of feedback strongly depends on the gas structure. It is most effective when the gas is newly collapsing and has a far smaller impact on dense regions.

\section*{Acknowledgments}

Numerical computations were carried out on Cray XC30 at the Center for Computational Astrophysics (CfCA) of the National Astronomical Observatory of Japan. The authors would like to thank the YT development team for many helpful analysis support \citep{Turk2011}. EJT was funded by the MEXT grant for the Tenure Track System. This work is partly supported by JSPS Grant-in-Aid for Scientific Research Number 15K0514.

\bsp

\label{lastpage}


\begin{thebibliography}{99}
\bibitem[Abel \& Wandelt(2002)]{ART1} Abel, T., \& Wandelt, B.~D.\ 2002, \mnras, 330, L53
\bibitem[Alves et al.(2001)]{Alves2001} Alves, J.~F., Lada, C.~J., \& Lada, E.~A.\ 2001, \nat, 409, 159 
\bibitem[Bate(2009)]{Bate2009} Bate, M.~R.\ 2009, \mnras, 392, 1363 
\bibitem[Benincasa et al.(2013)]{Benincasa2013} Benincasa, S.~M., Tasker, E.~J., Pudritz, R.~E., \& Wadsley, J.\ 2013, \apj, 776, 23
\bibitem[Bonnor(1956)]{Bonnor1956} Bonnor, W.~B.\ 1956, \mnras, 116, 351
\bibitem[Bryan et al.(2014)]{Enzo} Bryan, G.~L., Norman, M.~L., O'Shea, B.~W., et al.\ 2014, \apjs, 211, 19 
\bibitem[Cartwright \& Whitworth(2004)]{Q} Cartwright, A., \& Whitworth, A.~P.\ 2004, \mnras, 348, 589
\bibitem[Chabrier(2005)]{Chabrier2005} Chabrier, G.\ 2005, The Initial Mass Function 50 Years Later, 327, 41 
\bibitem[Dale et al.(2014)]{Dale2014} Dale, J.~E., Ngoumou, J., Ercolano, B., \& Bonnell, I.~A.\ 2014, \mnras, 442, 694
\bibitem[Dale et al.(2012)]{DaleMII} Dale, J.~E., Ercolano, B., \& Bonnell, I.~A.\ 2012, \mnras, 424, 377
\bibitem[Dale et al.(2012b)]{Dale2012b} Dale, J.~E., Ercolano, B., \& Bonnell, I.~A.\ 2012, \mnras, 427, 2852
\bibitem[Dobbs et al.(2015)]{Dobbs2015} Dobbs, C.~L., Pringle, J.~E., \& Duarte-Cabral, A.\ 2015, \mnras, 446, 3608 
\bibitem[Donovan Meyer et al.(2013)]{DonovanMeyer2013} Donovan Meyer, J., Koda, J., Momose, R., et al.\ 2013, \apj, 772, 107 
\bibitem[Federrath(2015)]{Federrath2015} Federrath, C.\ 2015, \mnras, 450, 4035 
\bibitem[Federrath et al.(2014)]{Federrath2014} Federrath, C., Schr{\"o}n, M., Banerjee, R., \& Klessen, R.~S.\ 2014, \apj, 790, 128
\bibitem[Ferland et al.(1998)]{Ferland1998} Ferland, G.~J., Korista, K.~T., Verner, D.~A., et al.\ 1998, \pasp, 110, 761 
\bibitem[Fujimoto et al.(2014)]{Fujimoto2014} Fujimoto, Y., Tasker, E.~J., Wakayama, M., \& Habe, A.\ 2014, \mnras, 439, 936 
\bibitem[Ginsburg et al.(2012)]{GMC1} Ginsburg, A., Bressert, E., Bally, J., \& Battersby, C.\ 2012, \apjl, 758, L29
\bibitem[G{\'o}rski et al.(2005)]{HEALPIX} G{\'o}rski, K.~M., Hivon, E., Banday, A.~J., et al.\ 2005, \apj, 622, 759
\bibitem[Harper-Clark \& Murray(2009)]{HarperClark2009} Harper-Clark, E., \& Murray, N.\ 2009, \apj, 693, 1696 
\bibitem[Hughes et al.(2013)]{Hughes2013} Hughes, A., Meidt, S.~E., Colombo, D., et al.\ 2013,  ApJ, 779, 46 
\bibitem[Kainulainen et al.(2014)]{Kainulainen2014} Kainulainen, J., Federrath, C., \& Henning, T.\ 2014, Science, 344, 183 
\bibitem[Koenig et al.(2012)]{Koenig2012} Koenig, X.~P., Leisawitz, D.~T., Benford, D.~J., et al.\ 2012, \apj, 744, 130 
\bibitem[Krumholz et al.(2014)]{Krumholz2014} Krumholz, M.~R., Bate, M.~R., Arce, H.~G., et al.\ 2014, Protostars and Planets VI, 243
\bibitem[Krumholz et al.(2010)]{Krumholz2010} Krumholz, M.~R., Cunningham, A.~J., Klein, R.~I., \& McKee, C.~F.\ 2010, \apj, 713, 1120 
\bibitem[Krumholz \& Matzner(2009)]{Krumholz2009} Krumholz, M.~R., \& Matzner, C.~D.\ 2009, \apj, 703, 1352
\bibitem[Krumholz \& Tan(2007)]{Krumholz2007} Krumholz, M.~R., \& Tan, J.~C.\ 2007, \apj, 654, 304
\bibitem[Lada et al.(2010)]{GMC2} Lada, C.~J., Lombardi, M., \& Alves, J.~F.\ 2010, \apj, 724, 687
\bibitem[Larson(1981)]{Larson1981} Larson, R.~B.\ 1981, MNRAS, 194, 809 
\bibitem[Mac Low et al.(1998)]{MacLow1998} Mac Low, M.-M., Klessen, R.~S., Burkert, A., \& Smith, M.~D.\ 1998, Physical Review Letters, 80, 2754 
\bibitem[McKee \& Williams(1997)]{McKee1997} McKee, C.~F., \& Williams, J.~P.\ 1997, \apj, 476, 144 
\bibitem[Meidt et al.(2015)]{Meidt2015} Meidt, S.~E., Hughes, A., Dobbs, C.~L., et al.\ 2015, \apj, 806, 72 
\bibitem[Murray(2011)]{Murray2011} Murray, N.\ 2011, \apj, 729, 133 
\bibitem[Offner \& Krumholz(2009)]{Offner2009a} Offner, S.~S.~R., \& Krumholz, M.~R.\ 2009, \apj, 693, 914 
\bibitem[Offner et al.(2009)]{Offner2009b} Offner, S.~S.~R., Klein, R.~I., McKee, C.~F., \& Krumholz, M.~R.\ 2009, \apj, 703, 131 
\bibitem[Rey-Raposo et al.(2015)]{ReyRaposo2015} Rey-Raposo, R., Dobbs, C., \& Duarte-Cabral, A.\ 2015, \mnras, 446, L46
\bibitem[Rogers \& Pittard(2013)]{Rogers2013} Rogers, H., \& Pittard, J.~M.\ 2013, \mnras, 431, 1337 
\bibitem[S{\'a}nchez \& Alfaro(2010)]{Q-parameter} S{\'a}nchez, N., \& Alfaro, E.~J.\ 2010, Star Clusters: Basic Galactic Building Blocks Throughout Time and Space, 266, 524 
\bibitem[Schaerer(2003)]{Schaerer2003} Schaerer, D.\ 2003, \aap, 397, 527
\bibitem[Stone \& Norman(1992)]{Zeus} Stone, J.~M., \& Norman, M.~L.\ 1992, \apjs, 80, 753
\bibitem[Urban et al.(2010)]{Urban2010} Urban, A., Martel, H., \& Evans, N.~J., II 2010, \apj, 710, 1343
\bibitem[Smith et al.(2008)]{CLOUDY} Smith, B., Sigurdsson, S., \& Abel, T.\ 2008, \mnras, 385, 1443
\bibitem[Tasker \& Tan(2009)]{Tasker2009} Tasker, E.~J., \& Tan, J.~C.\ 2009, \apj, 700, 358 
\bibitem[Tasker et al.(2015)]{Tasker2015} Tasker, E.~J., Wadsley, J., \& Pudritz, R.\ 2015, \apj, 801, 33 
\bibitem[Turk et al.(2011)]{Turk2011} Turk, M.~J., Smith, B.~D., Oishi, J.~S., et al.\ 2011, \apjs, 192, 9 
\bibitem[Truelove et al.(1997)]{Truelove1997} Truelove, J.~K., Klein, R.~I., McKee, C.~F., et al.\ 1997, \apjl, 489, L179
\bibitem[Walch et al.(2013)]{Walch2013} Walch, S., Whitworth, A.~P., Bisbas, T.~G., W{\"u}nsch, R., \& Hubber, D.~A.\ 2013, \mnras, 435, 917 
\bibitem[Whitworth et al.(1994)]{Whitworth1994} Whitworth, A.~P., Bhattal, A.~S., Chapman, S.~J., Disney, M.~J., \& Turner, J.~A.\ 1994, \aap, 290, 421 
\bibitem[Williamson et al.(2014)]{Williamson2014} Williamson, D.~J., Thacker, R.~J., Wurster, J., \& Gibson, B.~K.\ 2014, \mnras, 442, 3674 
\bibitem[Wise \& Cen(2009)]{Wise2009} Wise, J.~H., \& Cen, R.\ 2009, \apj, 693, 984
\bibitem[Wise \& Abel(2011)]{ART2} Wise, J.~H., \& Abel, T.\ 2011, \mnras, 414, 3458
\bibitem[Wise et al.(2012)]{popII} Wise, J.~H., Abel, T., Turk, M.~J., Norman, M.~L., \& Smith, B.~D.\ 2012, \mnras, 427, 311
\bibitem[W{\"u}nsch et al.(2010)]{Wunsch2010} W{\"u}nsch, R., Dale, J.~E., Palou{\v s}, J., \& Whitworth, A.~P.\ 2010, \mnras, 407, 1963 




\end{thebibliography}
\end{document}